\documentclass[conference,10pt]{IEEEtran}

\usepackage{amssymb,amsmath,amsfonts,bm,epsfig,graphicx,theorem,latexsym}
\usepackage{rotating,setspace,latexsym,epsf,color,subfigure}
\usepackage{cite,float,epsf}
\newtheorem{Theorem}{Theorem}

\newenvironment{Proof}[1]{\medskip\par\noindent
{\bf Proof:\,}\,#1}{{\mbox{\,$\blacksquare$}\par}}

\def \n2{{N_0 \over 2}}

\def \h5{\hspace{0.5in}}

\begin{document}

%\IEEEoverridecommandlockouts
%\pagestyle{empty}

\title{Intermittent Status Updating Through Joint Scheduling of Sensing and Retransmissions\vspace{-0.1in}}

\author{\IEEEauthorblockN{Omur Ozel \qquad Parisa Rafiee}
\IEEEauthorblockA{Department of Electrical and Computer Engineering\\
George Washington University, Washington DC 20052\\
{\it ozel@gwu.edu \qquad rafiee@gwu.edu}}\vspace{-0.2in}} 

\maketitle 

\begin{abstract} 
Consider an energy harvesting node where generation of a status update message takes non-negligible time due to sensing, computing and analytics operations performed before making update transmissions. The node has to harmonize its (re)transmission strategy with the sensing/computing. We call this general set of problems \textit{intermittent status updating}. In this paper, we consider intermittent status updating through \textit{non-preemptive} sensing/computing (S/C) and transmission (Tx) operations, each costing a single energy recharge of the node, through an erasure channel with (a) perfect channel feedback and (b) no channel feedback. The S/C time for each update is independent with a general distribution. The Tx queue has a single data buffer to save the latest packet generated after the S/C operation and a single transmitter where transmission time is deterministic. Once energy is harvested, the node has to decide whether to activate S/C to generate a new update or to (re)send the existing update (if any) to the receiver. We prove that when feedback is available average peak age of information (AoI) at the receiver is minimized by a threshold-based policy that allows only \textit{young} packets to be (re)sent or else generates a new update. We additionally propose window based and probabilistic retransmission schemes for both cases (a) and (b) and obtain closed form average peak AoI expressions. Our numerical results show average peak AoI performance comparisons and improvements.    
\end{abstract}

%\newpage
%\pagestyle{plain}

%\setcounter{page}{1}
%\pagenumbering{arabic}

\section{Introduction}

Age of Information (AoI) measures staleness of available information at a monitoring receiver. Since the pioneering work in \cite{kaul2012real}, the AoI metric has been widely used for timely information updating problems; see \cite{yates2020age} for a recent comprehensive survey of AoI and its applications.

%\cite{bedewy2017age, talak2017minimizing, yates2018age, maatouk2018age,Alabbasi2018JointIF,Gong2019ReducingAF,xu2019peak}. 

In this paper, we consider an \textit{intermittent status updating} problem that extends the one introduced recently in \cite{ozel2020}. In particular, the problem entails an energy harvesting node sending update packets after a \textit{non-preemptive} sensing/computing (S/C) operation and subsequent transmission (Tx) operation, each costing a single energy recharge of the node. This problem is motivated by intermittent computing based energy harvesting systems\footnote{We refer the reader to \cite{hester2017timely,lucia2017intermittent} for typical characteristics of intermittent computing.} where operations are incrementally performed one by one through charge/recharge cycles. Once power is restored after a recharge interval, the node has to decide whether to maintain forward progress or to start again with a fresh status update. The new aspects we will tackle in this paper are the presence of channel erasures and possible feedback about the sent update.

Energy harvesting communication systems have been a topic of extensive research interest in AoI analysis and optimization \cite{yates2015lazy,feng2018minimizing,arafa2019using,ceran2019reinforcement,krikidis2019average,rafiee2020active} where emphasis is mainly on average AoI with possible channel erasures and feedback under offline and online knowledge of energy harvests. See \cite{yates2020age} for a more complete list. A line of research that is closely related to the status updating in energy harvesting communication systems in \cite{ceran2019average,gu2019timely,gong2018energy} is on resource constraints and age-energy tradeoff. We also note references \cite{rafiee2020age,arafa2019timely} that connect to computation communication synergies with average cost considerations. Reference \cite{xu2019peak} is on peak AoI in data preprocessing based IoT networks. \cite{champati2020minimum} provides a general optimality of threshold based schemes to preempt service. On another related research front, \cite{costa2016age} introduces packet management to improve the average AoI at the monitoring node. Reference \cite{kam2018age} considers packet deadlines to discard outdated packets for data freshness.  

\begin{figure}[!t]
\centering{
\hspace{-0.0cm} 
\includegraphics[totalheight=0.19\textheight]{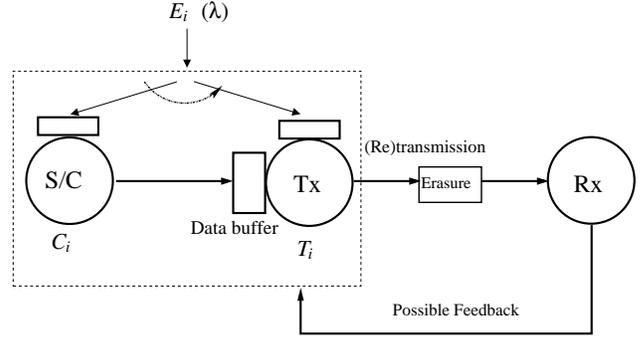}}\vspace{-0.1in}
\caption{\sl System model for an EH node performing S/C, Tx and ReTx operations through an erasure channel using harvested energy with possible feedback.}\vspace{-0.15in}
\label{fig:1} 
\end{figure}

In our current work, we build on the intermittent status updating model in \cite{ozel2020} with added new aspects due to channel erasures and availability of feedback about sent update in an energy harvesting node as in Fig. \ref{fig:1}. In our model, the time it takes for S/C operation is an independent random variable with a general distribution, and transmission time is deterministic. A single unit data buffer is available for storing the latest packet during recharge interval (also referred as power outage time). \textit{S/C operation can not be preempted.} Once energy recharge is maintained, the node allocates the whole energy for S/C, Tx or reTx operations. The node may want to reTx a generated update if it is not successfully delivered in the presence of feedback or to increase the chances of success in the absence of feedback. This retransmission attempt stops when success occurs if channel feedback is present or is simply preempted due to the time spent on sending a given update. We prove that when feedback is available average peak age of information (AoI) at the receiver is minimized by a threshold-based stopping rule that rejects old packets in the Tx buffer and keeps on retransmitting until a threshold or successful delivery. Earlier papers \cite{gu2019timely,gong2018energy} consider similar retransmission schemes under sensing costs. Our results differ from them due to the new intermittent status updating model with dynamic energy levels and our exclusive focus on the average peak AoI with explicit optimality proof in the presence of feedback. We are interested in the case when there is partial or no channel feedback available. Our results provide an opportunity to gauge optimal average peak AoI performance improvement gained by channel feedback and we plan to shape our future work toward this direction. 

The rest of the paper will flow as follows: In Section II, we cover details of the model. In Section III, we present our main result as well as window based and probabilistic retransmission schemes with feedback. In Section IV, we cover window based and probabilistic retransmission schemes without feedback. We present numerical results in Section V and conclude our work in Section VI.

\section{The Model}
\label{sec:Model}

We consider an intermittent computing based energy harvesting node shown in Fig. \ref{fig:1}. S/C represents the initial operation to generate a status update packet while Tx is transmission. We assume that there is always a packet to generate and it takes non-negligible time to complete generation through the S/C operation, which is assumed to be nonpreemptive. Generated packet starts aging once sensing is activated and then an update packet enters the transmission queue. There is a single data buffer to save the latest arriving packet when the system is in power outage (i.e., recharging). The channel through which updates travel is an erasure channel.

The transmitter has three choices to make once it recharges energy: (i) generate a new update through an S/C operation, which naturally discards the data available; (ii) send an update through a Tx operation, and (iii) resend an earlier generated update through another Tx (or reTx) operation. Once Tx or reTx is completed, the receiver (Rx) has the most recent update with probability $(1-p_e)$ where $p_e$ is the erasure probability of the channel. Throughout the document, $0\leq p_e <1$ is assumed. Energy is depleted at the end of S/C and (re)Tx operations. Time spent in S/C for update generation is independent with a general distribution $f_C(c)$, $c \geq 0$ with well-defined mean $\mathbb{E}[C]$. The time $C$ could be viewed as the cumulative time that potentially takes multiple charge/recharge cycles. Transmission time is a deterministic number $D$ known to the transmitter and receiver. That is, sent update is received with probability $(1-p_e)$ after $D$ time units. During this time, it is not possible to recharge energy and generate a new update. 

The energy arrives to the EH node incrementally according to a Poisson process of rate $\lambda$. Consequently, time spent in power outage (or the time to recharge) is independent exponentially distributed with the same rate. We denote the time in outage as $I$. Once energy arrives, it is used to activate \textit{either} S/C \textit{or} Tx. If there is no update, energy is used for a new generation. At the end of an operation, the node is left with no energy. Upon new energy recharge, the node decides whether to (re)transmit an update (if any) or to generate one. 

\begin{figure}[!t]
\centering{
\hspace{-0.0cm} 
\includegraphics[totalheight=0.22\textheight]{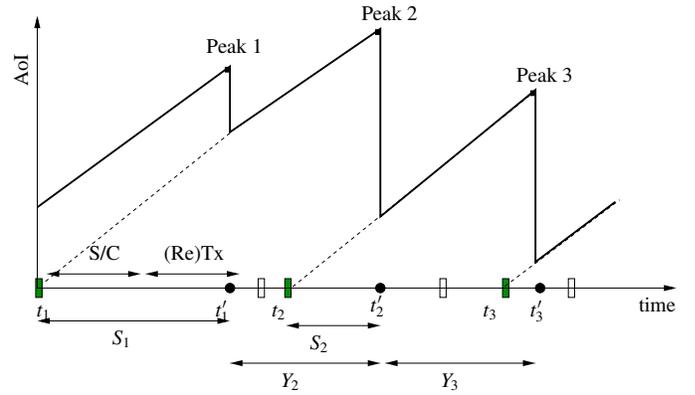}}\vspace{-0.15in}
\caption{\sl Example AoI evolution for general transmission schemes.}
\label{fig:2} 
\vspace{-0.1in}
\end{figure}

We use $t_i$ to denote the update $i$'s start of generation through sensing/computing, and $t_i'$ to denote the time stamp of the event that the update $i$ is received successfully. Age of Information (AoI) of update $i$ is the difference of the current time and the time stamp of the latest update at the receiver: 
\begin{align}
\Delta(t)=t-u(t)
\end{align}
$u(t)$ is the time stamp of the latest received packet at time $t$. The sample AoI evolution curve in Fig. \ref{fig:2} shows that update 1 starts aging in the S/C server at time $0$ when S/C starts. Once it is finished and the subsequent energy arrives, the node decides favorably to take the packet to the Tx data buffer right away and after possible Tx and reTx attempts its service ends at $t_1'$ when receiver successfully receives the sent update. Then, next energy is used to activate S/C to get the next update and the node decides to drop the update (shown as a hollow rectangle in Fig. \ref{fig:2}) after potential trials to reTx the generated update. The next energy arrival restarts the S/C operation and this time the node moves forward with transmission and the update is successfully received at $t_2'$.  

We define $Y_i$ as the length of time between the departures of updates $i-1$ and $i$ and $S_i$ as the system time for packet $i$ as shown in Fig. \ref{fig:2}. The average peak age of information is:
\begin{align}\label{paoi}
\mathbb{E}[\Delta^{p}]=\lim_{t \rightarrow \infty} \frac{1}{N_t} \sum_{i=1}^{N_t} (S_i + Y_{i+1}) 
\end{align}
Here, $N_t$ is the number of sent updates by time $t$. Our underlying assumption is that $N_t \rightarrow \infty$ and the system is ergodic in that the decisions to transmit, retransmit or generate an update are well behaving to allow existence of above limit. 

\section{Average Peak AoI Minimization with Feedback}

In this section, we consider average peak AoI minimization when strictly causal feedback about the success of transmissions is available. In this case, the transmission policy is designed with the knowledge of the success or failure of sent update. If an update is known to be successful, there is no need to resend that update and the node uses the next energy arrival for generating a new update. If, on the other hand, the update transmission is known to fail, the node has to decide whether to generate a new update or resend it once new energy arrives. We denote the channel feedback by $h_i$: If $h_i=1$, transmission is successful and $h_i=0$ otherwise. 

Our focus is on \textbf{renewal type Tx policies} where decision to (re)transmit or discard an update is taken based on observed time and events starting from the last successful transmission with randomization allowed. Thus, decisions after $t_i'$ are assumed independent of the past before $t_{i}'$. The same randomized policy is applied after $t_i'$ irrespective of the index $i$. Indeed, from our development, it will be clear that any dependence on the past history does not promise average peak AoI performance improvement when feedback is available. 

In view of the renewal policies, we will work with the generic variables for inter-departure time $Y$ and system time $S$. We express $Y$ and $S$ in this case as:
\begin{align}\label{bir}
Y &= I_0 + \sum_{k=1}^{\tilde{n}}  C_k + I_k + T_k \\
S &= C_{\tilde{n}} + I_{\tilde{n}} + T_{\tilde{n}} \label{iki}
\end{align} 
where $\tilde{n}$ is a stopping time with respect to $\{C_i+I_i + T_i\}$ and possibly $I_0$; that is, whether $\tilde{n}=n$ depends only on the realizations of $\{C_i+I_i + T_i\}_{i=1}^n$ and $I_0$. Here, $I_0$ represents the initial outage time experienced after a successful update delivery and $I_i$, $i \geq 1$ are outage times experienced after the S/C operation. Note that $I_i$, $i\geq 0$ are independent exponentially distributed. $T_i$ is the time spent to send and resend the update $i$ including transmission times $D$ and time waited for new energy to arrive. As a matter of fact, we have 
\begin{equation*}
T_i=\left\{\begin{array}{ll}
\widehat{n}_iD + \sum_{j=1}^{\widehat{n}_i} I_{ji}, & \mbox{$h_{n_ii}=0$}\\
\widehat{n}_iD + \sum_{j=1}^{\widehat{n}_i - 1} I_{ji}, & \mbox{$h_{n_ii}=1$}
\end{array}
\right.
\end{equation*}
where $\widehat{n}_k$ denotes the retransmission index, is naturally conditioned on $C_k + I_k$, and is a stopping time with respect to $\{I_{jk}\}$, the energy recharge times for each Tx attempt $j$ for update $k$, and the channel feedback $\{h_{jk}\}$. Tx attempt $j$ for update $k$ is decided based on the strictly causal past channel feedback $\{h_{mk}\}$, $m<j$ and causal recharge times $\{I_{mk}\}$, $m\leq j$. If the node decides not to transmit at all, then $\widehat{n}_k=0$ and hence $T_k=0$. With probability one, $\widehat{n}_k=j^*$ if $h_{j^*k}=1$ and in this case $T_k$ includes only a transmission time $D$ for the Tx attempt $j^*$. On the other hand, if $h_{jk}=0$, transmission fails and whether $\widehat{n}_k=j$ or not is decided by the adopted Tx policy after the energy recharge following the failed transmission. Additionally, the outer stopping time is $\tilde{n}=k^*$ if and only if $h_{j^*k^*}=1$. Once update delivery is successful (i.e., when $h_{j^*k^*}=1$), the renewal cycle is reset and everything starts anew with the new energy recharge $I_0$. We may allow $\widehat{n}_k$ to depend on past history; however, it does not promise any improvement. This will be clear in the proof of our main result.

We call $C_i + I_i + T_i \triangleq A_i$ and $\tilde{n}$ is conditioned on natural filtrations $\mathcal{F}^{A}_i, i >0$ with $\mathcal{F}^{A}_i=\sigma(A_k, k\leq i)$ of the sequence $\{A_i\}$. We do not assume dependence on $I_0$ as it does not improve peak AoI; c.f. \cite{ozel2020}. Then, the average peak AoI is: $\mathbb{E}[I_0 + \sum_{k=1}^{\tilde{n}} A_k] + \mathbb{E}[A_{\tilde{n}}]$. We solve the following problem:
\begin{align} \label{prob}
\inf_{\widehat{n} \in \mathcal{M}} \mathbb{E}[\Delta^p]
\end{align}
where $\mathcal{M}$ is the space of all stopping times $\widehat{n}_k$ with well-defined mean values. This problem is in the form of a Markov Decision Process (MDP) with the state definition $(n,x)$. Here, $\tilde{n}=n$ represents the update index. Additionally, $C_i+I_i + \widehat{r}D +\sum_{j=1}^{\widehat{r}} I_{ji}=x$, with $\widehat{r}$ being the retransmission index, represents the age of the current update before making decision to Tx, reTx (same action in different $\widehat{r}$) or abandon it to generate a new update. Hence, action space is binary: (i) transmit existing update and (ii) abandon to generate a new update. We next state our main result in the following theorem: 
\begin{Theorem}\label{thm1}
When feedback is available, optimal rule that minimizes average peak AoI $\mathbb{E}[\Delta^p]$ is to transmit when $A_i < W_{th}$ until successful delivery and to stop (or never start) transmitting when $A_i \geq W_{th}$ where $W_{th}$ is a finite threshold. The optimal average peak AoI in terms of this threshold is:
\[\frac{D}{1-p_e} + W_{th}+D + \frac{1}{(1-p_e)\lambda}\] 
\end{Theorem}
\begin{Proof} We define the cost to go $g(x)$ as follows: \[g(x) \triangleq \inf_{\widehat{n}_x \in \mathcal{M}}  \mathbb{E}[\Delta^p | C_1 + I_1 = x] \]
where $\widehat{n}_x$ refers to all stopping times with respect to the ensuing energy recharge times for each Tx attempt and following channel feedback conditioned on the initial state $C_1+I_1 = x$.  Let us first decompose $g(x)$ as follows:
\begin{align*}
g(x) = \mathbb{E}[I_0] + x + s(x)
\end{align*}
where we define \[s(x) \triangleq \inf_{\widehat{n}_x \in \mathcal{M}}\mathbb{E}[T_1+\sum_{k=2}^{\tilde{n}} A_k] + \mathbb{E}[A_{\tilde{n}}]\] and dependence on $x$ is buried in the stopping times $\widehat{n}_k$ in $T_k$. We additionally define $f_{opt}$ as part of the optimal cost: \[\mathbb{E}[g(C+I)] = \mathbb{E}[I_0] + f_{opt} = \frac{1}{\lambda} + f_{opt}\]

Once the node observes $C_1+I_1=x$, it has two choices: To transmit or to abandon and generate a new update. If transmission is chosen, with probability $1-p_e$ it is successful, $\tilde{n}=1$, $\widehat{n}_1=1$, $A_1=x+D$ and the final cost is $\frac{1}{\lambda}+2(x+D)$, the cycle ends. With probability $p_e$ it fails, the state is carried to $x+D+I$ where $I$ is time spent in power outage and since we are still deciding on update 1, $n$ remains at 1. In this case, the average cost is $\frac{1}{\lambda}+\mathbb{E}[x+D+I + s(x+D+I)]=\frac{1}{\lambda}+x+D+\frac{1}{\lambda}+\mathbb{E}[s(x+D+I)]$. So, if transmission is chosen at state $x$, the average cost is \[\frac{1}{\lambda}+x+D + (1-p_e)(x+D) + \frac{p_e}{\lambda} + p_e\mathbb{E}[s(x+D+I)]\] On the other hand, if generating a new update is chosen, the decision process makes a new start, the outer index is carried from 1 to 2 and the average cost paid is \[\frac{1}{\lambda} +x+f_{opt}\] Therefore, for all $x$ that satisfies the following inequality:  \[\mbox{\textbf{(i)}:} \ \mbox{$D + (1-p_e)(x+D) + \frac{p_e}{\lambda} + p_e\mathbb{E}[s(x+D+I)] \geq f_{opt}$}\] optimal choice is to generate a new update and we have $s(x)=f_{opt}$. Similarly, for all $x \geq 0$ that satisfies the following: \[\mbox{\textbf{(ii)}:} \ \mbox{$D + (1-p_e)(x+D) + \frac{p_e}{\lambda} + p_e\mathbb{E}[s(x+D+I)] < f_{opt}$}\] optimal choice is to transmit and we have \[s(x)=D + (1-p_e)(x+D) + \frac{p_e}{\lambda} + p_e\mathbb{E}[s(x+D+I)]\]
We claim that if for some $x$, $s(x)=f_{opt}$, it holds for all larger $x$. Indeed, it is clear that for \[ x \geq W_{ub} \triangleq \frac{f_{opt} - \frac{p_e}{\lambda} - (2-p_e)D}{1-p_e}\] inequality \textbf{(i)} is satisfied and $s(x)=f_{opt}$. On the other hand, for any $x$ satisfying \[x < W_{th} \triangleq \frac{(1-p_e)f_{opt} - \frac{p_e}{\lambda} - (2-p_e)D}{1-p_e}\] the inequality \textbf{(ii)} is satisfied and $s(x)<f_{opt}$. Note that $W_{th} < W_{ub}$. Also note that $x=W_{th}$ satisfies inequality \textbf{(ii)} with equality when $\mathbb{E}[s(x+D+I)]=f_{opt}$. Finally, we observe that for any $\delta \geq 0$ and $x=W_{ub}-\delta$ that satisfy inequality \textbf{(i)}, we have $s(x)=f_{opt}$ and $\mathbb{E}[s(x+D+I)] = f_{opt}$ for all $x \geq W_{ub} - \delta$. This, in turn, means that for any $\delta \geq 0$ satisfying \[D + (1-p_e)(W_{ub} - \delta+D) + \frac{p_e}{\lambda} + p_ef_{opt} \geq f_{opt}\]
$s(x) = f_{opt}$ holds. We conclude that the inequality \textbf{(i)} is satisfied by all $x \geq W_{th}$ where $W_{th}=\inf\{x: s(x)=f_{opt}\}$ and $W_{th}$ satisfies:
\begin{align*}
D + (1-p_e)(W_{th}+D) + \frac{p_e}{\lambda} = (1-p_e)f_{opt}
\end{align*} 
This relation is sufficient to assert the expression of optimal average peak AoI, which is $\frac{1}{\lambda}+f_{opt}$, in terms of $W_{th}$.

To conclude, we let the node observe $n$ and $A_n=x$ after an energy recharge. Also assume $I_0$ and past $A_{k<n}$ are given. If the node decides to transmit, the average cost paid is 
\begin{align*} 
I_0 + \sum_{j=1}^{n-1}A_j + x+D+&(1-p_e)(x+D) \\ &+ p_e\mathbb{E}[I + s(x+I+D)] 
\end{align*} 
If, on the other hand, the node decides to generate a new update, then on average it has to pay the cost
\begin{align*}
I_0 +  \sum_{j=1}^{n-1}A_j + x + f_{opt}
\end{align*}
Then, due to optimal stopping criterion in discrete time (see e.g. \cite{shiryaev2007optimal}), the optimal stopping time is obtained by comparing the two costs: If $D+(1-p_e)(x+D) + p_e\mathbb{E}[I + s(x+I+D)] < f_{opt}$, Tx is chosen and otherwise generating a new update is chosen. We conclude that when $A_i$ hits $A_i=x$ for $x$ that satisfies $x \leq W_{th}$, then, existing update must be sent and resent until $A_i$ goes above $W_{th}$ or success is achieved. 
\end{Proof}

%We next consider two benchmark schemes and obtain closed form expressions for average peak AoI.

\subsection{Window Based Retransmission}

In this subsection, we consider a window based retransmission scheme. The scheme is defined by two parameters: $W_s$ for threshold and $B$ for maximum number of retransmissions. When $C_i+I_i < W_s$, Tx commits to transmit the update $i$ and at most $B$ times retransmissions are allowed for successful transmission. If success is achieved, then transmission stops and a new update is generated. If no success is observed, then a new update is generated at the end of $B$ retransmissions. 

We first note that in this case $T_k$ are i.i.d. in view of the identical distribution of inner stopping times $\widehat{n}_k$ and so are $A_k$. In particular, $\widehat{n}_k$ are independent from $I_i$ and distributed as geometric with success probability $1-p_e$ truncated at $B$: 
\begin{equation*}
\mbox{Pr}(\widehat{n}_k=v)=\left\{\begin{array}{ll}
(1-p_e)p_e^{v-1}, & \mbox{$1\leq v \leq B$}\\
p_e^{B}, & \mbox{$v=B^+$}
\end{array}
\right.
\end{equation*}
where $B$ and $B^+$ are taken as two different symbols with the same value equal to B. This is to differentiate between success and failure at the end of the $B$th transmission. Note also that $A_k$ are independent of $I_0$. By Wald's identity \cite{gallager2012discrete}, we have the following:
$\mathbb{E}[\sum_{k=1}^{\tilde{n}} A_k] = \mathbb{E}[\tilde{n}]\mathbb{E}[A]$ where the outer stopping time $\tilde{n}$ increases as long as $C_i+I_i>W_s$ or $\widehat{n}_k=B^+$, which happens with probability $p_e^B+\mbox{Pr}(C_i+I_i>W_s) - p_e^B\mbox{Pr}(C_i+I_i>W_s)$. Stopping happens in the complement of this event. Hence, we have \[\mathbb{E}[\tilde{n}] = \frac{1}{(1-p_e^B)\mbox{Pr}(C_i+I_i \leq W_s)} \]
In this case, expectation of $A$ is: $\mathbb{E}[A]=\mathbb{E}[C] + \frac{1}{\lambda}+ \mathbb{E}[T]$
where 
\begin{align*}
\mathbb{E}[T]=&\mbox{Pr}(C_i+I_i > W_s)\mathbb{E}[T|C_i+I_i > W_s] \\&\qquad + \mbox{Pr}(C_i+I_i \leq W_s)\mathbb{E}[T|C_i+I_i \leq W_s] \\ =&\mbox{Pr}(C_i+I_i \leq W_s)p_e^B \mathbb{E}[T|\widehat{n}= B^+]\\&+\mbox{Pr}(C_i+I_i \leq W_s)(1-p_e^B)\mathbb{E}[T|\widehat{n}\neq B^+]
\end{align*}
where $\mathbb{E}[T|C_i+I_i > W_s]=0$ since no transmission occurs if initial age is larger than $W_s$. Note that $\mathbb{E}[T|\widehat{n}= B^+]=B(D+\frac{1}{\lambda})$. We also calculate 
\begin{align*}
(1-p_e^B)\mathbb{E}[T|\widehat{n}\neq &B^+]=(\frac{1-p_e^{B}}{1-p_e}-Bp_e^{B})D\\&+(\frac{1-p_e^{B}}{1-p_e}-(1-p_e^B)-Bp_e^{B})\frac{1}{\lambda}
\end{align*}
Finally, note that the stopped mean is:
\begin{align*}
\mathbb{E}[A_{\tilde{n}}]&=\mathbb{E}[C_{\tilde{n}}+I_{\tilde{n}}+T_{\tilde{n}}] \\ &= \mathbb{E}[C+I | C+I<W_s] + \mathbb{E}[T|\widehat{n}\neq B^+] 
\end{align*}
We therefore have the average peak AoI
\begin{align*}
\mathbb{E}[\Delta^p] &= \frac{1}{\lambda}+\mathbb{E}[\tilde{n}]\mathbb{E}[A] + \mathbb{E}[A_{\tilde{n}}] 
\end{align*}

\subsection{Probabilistic Retransmission} 

In this subsection, we consider a probabilistic retransmission scheme. The scheme is defined by two parameters: $W_p$ for threshold and $p_{Tx}$ for transmission probability. When $C_i+I_i < W_p$, the transmitter commits to transmit with probability $p_{Tx}$ and generates a new update with probability $1-p_{Tx}$.

In this case, the inner stopping time $\widehat{n}=\min\{F_1,F_2\}$ where $F_1,F_2$ have independent geometric distributions with success probabilities $(1-p_e)$ and $1-p_{Tx}$, respectively. In particular, we let $F_2$ start from $0$ while $F_1$ starts from 1. If $F_1\leq F_2$, the attempted update delivery is successful and otherwise it is abandoned without a successful delivery.  Note also that $\widehat{n}$ is independent from recharge times $I_i$. Therefore, $A_i$ are i.i.d. and by Wald's identity:
$\mathbb{E}[\sum_{k=1}^{\tilde{n}} A_k] = \mathbb{E}[\tilde{n}]\mathbb{E}[A]$ where the outer stopping time $\tilde{n}$ stops when $C_i+I_i \leq W_p$ and $\widehat{n}=F_1$, with probability $\mbox{Pr}(C_i+I_i \leq W_p)\frac{p_{Tx}(1-p_e)}{1-p_ep_{Tx}}$ since the probability of $F_1 \leq F_2$ is $\frac{p_{Tx}(1-p_e)}{1-p_ep_{Tx}}$. We get: \[\mathbb{E}[\tilde{n}] = \frac{1-p_ep_{Tx}}{\mbox{Pr}(C_i+I_i \leq W_p)p_{Tx}(1-p_e)} \]
In this case, conditional expectation of $T$ is:
\begin{align*}
\mathbb{E}[T|C+I\leq W_p]=& \frac{(1-p_{Tx})p_ep_{Tx}}{(1-p_ep_{Tx})^2} (D+\frac{1}{\lambda})\\ &+\frac{p_{Tx}(1-p_e)}{(1-p_ep_{Tx})^2}(D+p_ep_{Tx}\frac{1}{\lambda})
\end{align*}
To see this, we note that the event $\widehat{n}=k$ happens in two mutually exclusive ways: First is when $k$ transmissions are made and all of them are erased and the node decides to generate a new update on the $k+1$st time. This happens with probability $\mbox{Pr}(\widehat{n}=k,1)=p_e^kp_{Tx}^k(1-p_{Tx})$, $k=0,1,...$. Second is when $k$ transmissions are made, they are erased until $k-1$st trial and the $k$th one is successful. This happens with probability $\mbox{Pr}(\widehat{n}=k,2)=(1-p_e)p_e^{k-1}p_{Tx}^k$, $k=1,2,...$. We then write
\begin{align*}
\mathbb{E}[T|C+I\leq W_p] = &\sum_{k=0}^{\infty} p_e^kp_{Tx}^k(1-p_{Tx})\mathbb{E}[T|\widehat{n}=k,1] \\ &+ \sum_{k=1}^{\infty} (1-p_e)p_e^{k-1}p_{Tx}^k\mathbb{E}[T|\widehat{n}=k,2] \\ =&\sum_{k=0}^{\infty} p_e^kp_{Tx}^k(1-p_{Tx})k(D+\frac{1}{\lambda}) \\ &+ \sum_{k=1}^{\infty} (1-p_e)p_e^{k-1}p_{Tx}^k(kD+(k-1)\frac{1}{\lambda})
\end{align*}  
Each summation corresponds to each group of expressions for $\mathbb{E}[T|C+I\leq W_p]$. We then combine to get
\begin{align*}
\mathbb{E}[T]=&\mbox{Pr}(C_i+I_i \leq W_p)\mathbb{E}[T|C+I\leq W_p]\\&+\mbox{Pr}(C_i+I_i > W_p)\mathbb{E}[T|C+I> W_p] \\ = &\mbox{Pr}(C_i+I_i \leq W_p)\mathbb{E}[T|C+I\leq W_p]
\end{align*}
since $\mathbb{E}[T|C+I> W_p]=0$.
We obtain $\mathbb{E}[A]=\mathbb{E}[C] + \frac{1}{\lambda}+ \mathbb{E}[T]$.
We also note that the stopped mean is:
\begin{align*}
\mathbb{E}[A_{\tilde{n}}]&=\mathbb{E}[C_{\tilde{n}}+I_{\tilde{n}}+T_{\tilde{n}}] \\ &= \mathbb{E}[C+I | C+I<W_p] + \frac{D+p_ep_{Tx}\frac{1}{\lambda}}{1-p_ep_{Tx}}
\end{align*}
We therefore have all components of the average peak AoI.

\section{The Case without Feedback}
\label{sec:eval}

In the case without feedback, we express $Y$ and $S$ as:
\begin{align}\label{bir1}
Y &= I_0 + \sum_{k=1}^{\tilde{n}} C_k + I_k + T_k \\
S &= C_{\tilde{n}} + I_{\tilde{n}} + T_{\tilde{n}} + T_{ext} \label{iki2}
\end{align} 
where $T_{ext}$ is the extra time spent on resending an update through reTx after its successful reception in $T_{\tilde{n}}$ time units. We can deduce these expressions through inspecting Fig. \ref{fig:2}. The extra time $T_{ext}$ is due to the lack of channel feedback and it must be added to either inter-departure time $Y$ or system time $S$ but not to both since each peak is affected by this extra time only once, which is observed in Fig. \ref{fig:2}. During this extra time AoI keeps climbing and it adds to the next peak. We assume that retransmission policy is of renewal type, stops retransmitting a given update with probability one and the extra delay is finite in the mean. Other than the lack of feedback, the definition of stopping times $\tilde{n}$ and $\widehat{n}$ are identical to the counterparts we used in the previous section. Specifically, $\tilde{n}$ counts the number of generated updates until a successful update transmission is achieved and $\widehat{n}$ reflects the retransmission index for a given update albeit this time with no knowledge of success of previous Tx attempts. 

Unlike the case with feedback, when feedback is absent we will not provide optimality proof of any type of policy. In the problem without feedback, potentially all past energy recharge times and instances must be accounted for while deciding on whether to transmit or generate a new update. Instead of considering the optimization, we will provide closed form expressions for the average peak AoI under two types of retransmission policies we covered in the case with feedback.

\subsection{Window Based Retransmission}

For the window based retransmission scheme, when feedback is not available, the transmitter commits to retransmit exactly $B$ times irrespective of the success of transmissions. This is of course after making sure that the initial age $C_i+I_i$ is below the set threshold $W_s$. At the end of $B$ retransmissions, a new update is generated. Note that $\mathbb{E}[\sum_{k=1}^{\tilde{n}} C_k + I_k + T_k]$ remains unchanged with respect to the window based retransmission scheme with feedback. We just have to calculate the average system time $S$ in (\ref{iki2}); that is, $\mathbb{E}[A_{\tilde{n}}]$. Note that the $B$ transmissions take place irrespective of success and therefore $T_{ext}$ and $T_{\tilde{n}}$ sums up to $B$ retransmission times:
\begin{align*}
\mathbb{E}[A_{\tilde{n}}]&=\mathbb{E}[C_{\tilde{n}}+I_{\tilde{n}}+T_{\tilde{n}}] \\ &= \mathbb{E}[C+I | C+I<W_s] + BD+(B-1)\frac{1}{\lambda} 
\end{align*}
This way, we obtain the corresponding expression for the average peak AoI.

\subsection{Probabilistic Retransmission}

For the probabilistic retransmission scheme, when feedback is not present, the transmitter commits to transmit with probability $p_{Tx}$ irrespective of an earlier successful transmission of the same update. The set threshold $W_s$ again serves the purpose of constraining the initial age. Similar to the window based retransmission, we just have to recalculate $\mathbb{E}[A_{\tilde{n}}]$. Note that since the node keeps on transmitting with probability $p_{Tx}$, we observe that $T_{ext}$ is the additional count with success probability $1-p_{Tx}$ starting from 0 conditioned on a successful update delivery is observed earlier. Due to the memoryless property of the geometric counting, $T_{ext}$ is also geometric and we have:
\begin{align*}
\mathbb{E}[A_{\tilde{n}}]&=\mathbb{E}[C_{\tilde{n}}+I_{\tilde{n}}+T_{\tilde{n}}] \\ &= \mathbb{E}[C+I | C+I<W_s] + \frac{D+p_ep_{Tx}\frac{1}{\lambda}}{1-p_ep_{Tx}} \\ &\qquad \qquad \qquad \qquad + (D+\frac{1}{\lambda})\frac{p_{Tx}}{1-p_{Tx}}
\end{align*}
This enables us to obtain average peak AoI.

\subsection{Threshold Based Retransmission}

We will also include the threshold based scheme in our experimentation with no feedback. In this case, the transmitter commits to resend an update until its age hits a specified threshold $W_T$. If the initial age $C_i+I_i > W_T$, then no transmission occurs and a new update is generated. This scheme is close to the optimal scheme with feedback. The major difference in this case is that the node keeps on sending until the age threshold $W_T$ is hit whereas the transmission attempts stop in a successful delivery when feedback is available. We note that it is not immediately possible to evaluate the probabilities and expectations under threshold based scheme. We will therefore resort to simulations to evaluate performances in the presence and absence of feedback. The recursive characteristic of the scheme makes it amenable to implement using MATLAB.\vspace{-0.05in}

\section{Numerical Results}

In this section, we provide numerical results illustrating average peak AoI performances of various schemes with and without feedback. Following the earlier work in \cite{ozel2020}, we consider a binary valued S/C time $C_i \in \{m_1, m_2\}$ with probabilities $p_1, p_2>0$ and $\mathbb{E}[C]=p_1m_1+p_2m_2$, $\mathbb{E}[C^2]=p_1m^2_1+p_2m^2_2$. We will aim to test the effects of arrival rate $\lambda$, erasure probability $p_e$ and variability of S/C time in average peak AoI performance. To this end, we consider a set of S/C time distributions $f_C$ with fixed mean $\mathbb{E}[C]=5$. We fix $m_1=1$ and parametrize $m_2$ and $p_2$ for fixed mean $5$. Let the parameter be $\theta>0$ so that $m_2=10+\theta$ and $p_2=\frac{4}{9+\theta}$ that guarantees fixed mean $5$. The variance is $\mbox{Var}(C)=20 + 4\theta$. We also set $D=1$ throughout.

In all schemes we evaluate, we use closed form expressions when available to plot the average peak AoI numerically optimized over a large range of parameters. We also verified the correctness of those expressions through random experiment based evaluations in MATLAB run more than $5\times 10^6$ times. For threshold based schemes, closed form expressions are not available. We just performed random experiments to obtain their average peak AoI performances and we optimize over the threshold. In the case with feedback, Theorem 1 suggests a relation between the optimal average peak AoI and the best threshold. In all random experiments we run, we observed that this relation holds and that other threshold levels yield higher average peak AoI. For the case without feedback, we observed in the experiments that the average peak AoI is quasiconvex with respect to the set threshold (this is typically the case in the literature on AoI) and made sure its optimality in a large set of thresholds. We ultimately note that the schemes that we refer to as ``best" are those that we observed as the best achievable and that we do not claim optimality over those schemes.

Let us define the random variable $M \triangleq C+I$. Its density function is $f_M(x)=p_1\lambda e^{-\lambda(x-m_1)}u(x-m_1) + p_2\lambda e^{-\lambda (x-m_2)}u(x-m_2)$ for $x\geq 0$ where $u(.)$ is the unit step function. We get the cumulative distribution (cdf) of $M$: 
\begin{align*}\nonumber
\int_{m^*}^x f_M(\alpha)d\alpha &= p_1(1-e^{-\lambda (x-m_1)})u(x-m_1) \\ &\hspace{0.2in}+ p_2(1-e^{-\lambda (x-m_2)})u(x-m_2) \end{align*} 
Similarly, we have the following integral:
\begin{align*} \nonumber \hspace{-0.10in}
\int_{m^*}^x \alpha f_M(\alpha)d\alpha &= p_1(m_1 + \frac{1}{\lambda} - (x + \frac{1}{\lambda})e^{-\lambda (x-m_1)}) u(x-m_1) \\ &\hspace{-0.5in} + p_2(m_2 + \frac{1}{\lambda} - (x + \frac{1}{\lambda})e^{-\lambda (x-m_2)}) u(x-m_2)
\end{align*}
which we use along with the cdf to evaluate $\mathbb{E}[M|M<W_{th}]$.

\begin{figure}[!t]
\centering{
\hspace{-0.3cm} 
\includegraphics[totalheight=0.29\textheight]{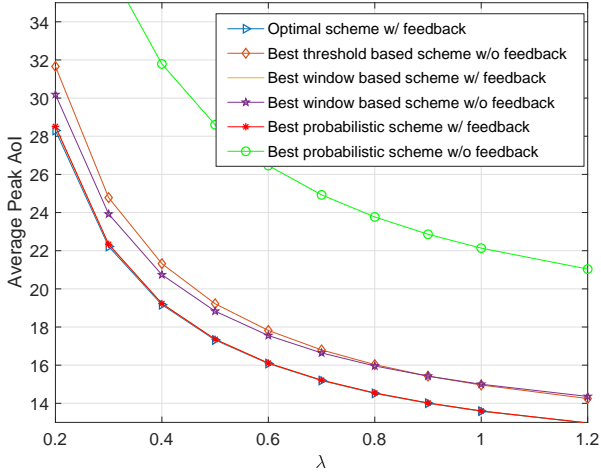}}\vspace{-0.15in}
\caption{\sl Average peak AoI with respect to energy arrival rate $\lambda$. }\vspace{-0.15in}
\label{numres:3} 
\end{figure}

We first consider average peak AoI vs $\lambda$. We plot in Fig. \ref{numres:3} the average peak AoI performances of all schemes when the parameter for $C$ is $\theta=10$ (i.e., $\mbox{Var}(C)=60$) and probability of error is $p_e=0.2$. We observe that when feedback is available the best window based and probabilistic schemes perform almost as good as the optimal threshold scheme, especially when $\lambda$ is large. The three schemes look coincident but there are minor differences observable when zoomed in. We also observe that when feedback is not available window based and threshold schemes have significantly smaller average peak AoI compared to the best probabilistic scheme. We will see this observation repeats in other plots. Between the threshold and the window based schemes, the latter outperforms the former for smaller $\lambda$ and the order changes for larger $\lambda$.

\begin{figure}[!t]
\centering{
\hspace{-0.3cm} 
\includegraphics[totalheight=0.29\textheight]{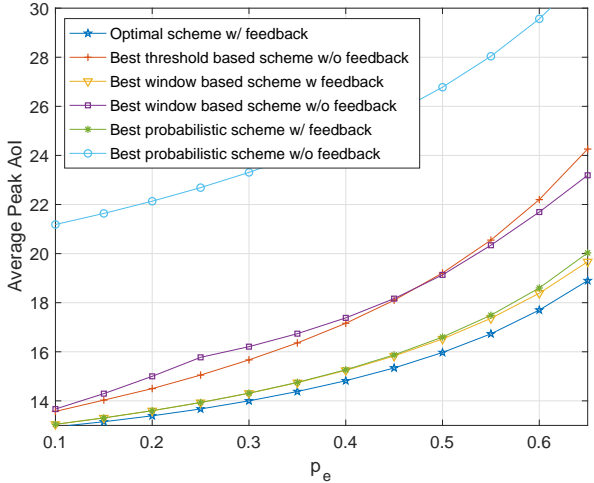}}\vspace{-0.15in}
\caption{\sl Average peak AoI with respect to erasure probability $p_e$. }\vspace{-0.15in}
\label{numres:2} 
\end{figure}

Next, we consider average peak AoI vs erasure probability in Fig. \ref{numres:2}. For this plot, we take $\theta=10$ (i.e., $\mbox{Var}(C)=60$) and $\lambda=1$. We observe that when feedback is present the optimal scheme significantly outperforms the other two schemes as $p_e$ is increased. This is understandable since the cost paid by committing to retransmit cannot meet the loss due to increased channel erasures. Therefore, limiting the retransmission attempts by a fixed threshold (known to be optimal) provides major improvement for larger $p_e$. The gap indeed grows for larger $p_e$ not shown in the plot. We also observe that when feedback is not present the best window based scheme outperforms the best threshold scheme for larger $p_e$. Nonsmooth behavior of the best window based scheme is due to the discreteness in the window size and abrupt change of the best window size as $p_e$ is varied.     

\begin{figure}[!t]
\centering{
\hspace{-0.3cm} 
\includegraphics[totalheight=0.29\textheight]{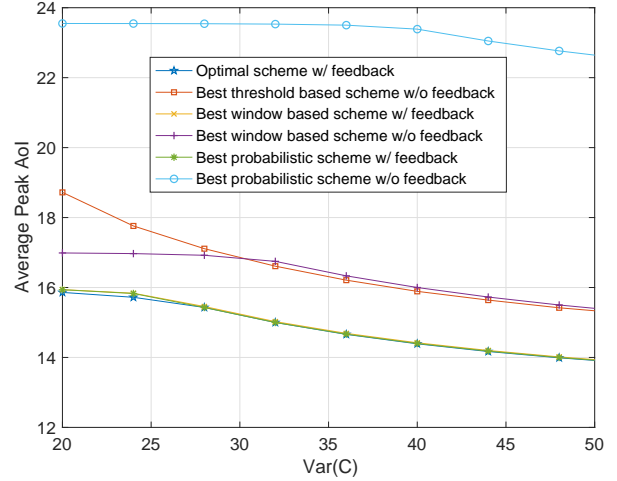}}\vspace{-0.15in}
\caption{\sl Average peak AoI with respect to variance of compute time $C$. }\vspace{-0.15in}
\label{numres:1} 
\end{figure}

Finally, we consider average peak AoI vs $\mbox{Var}(C)$ in Fig. \ref{numres:1}. We set $\lambda=1$, $p_e=0.2$ and vary $\theta$. We note that when feedback is available the three schemes perform almost equally well with an observable gap for small $\mbox{Var}(C)$ where the probabilistic scheme has some advantage. The ranking among the three schemes are similar to other plots when feedback is absent. We also observe the probabilistic scheme has the worst performance. We see by comparing the plots that $\lambda, p_e, \mbox{Var}(C)$ yield different orders among the three schemes depending on feedback availability. This calls for a more detailed study. \vspace{-0.05in}

\section{Conclusions}
\label{sec:conc}
In this paper, we consider an intermittent status updating problem where an energy harvesting node schedules S/C and retransmissions through an erasure channel with (a) perfect channel feedback and (b) no channel feedback. The node has S/C and Tx blocks and once energy is harvested, it decides whether to activate S/C to generate a new update or to (re)send the existing update (if any) to the receiver. We prove that in case (a) average peak AoI is minimized by a threshold-based policy. We also propose window based and probabilistic retransmission schemes in both cases and present numerical results that show average peak AoI performance comparisons.

\end{document}